\documentclass[prl,superscriptaddress,twocolumn,floatfix,showpacs,letterpaper]{revtex4}
\usepackage{amsmath}
\usepackage{graphicx}
\usepackage{times}
\usepackage{color}
\usepackage[normalem]{ulem}
\newcommand{\beq}{\begin{equation}}
\newcommand{\eeq}{\end{equation}}
\newcommand{\bea}{\begin{eqnarray}}
\newcommand{\eea}{\end{eqnarray}}

\newcommand{\RR}{{\bf R}}
\newcommand{\RRp}{{\bf R'}}
\newcommand{\PP}{{\mathcal{P}}}
\newcommand{\dd}{{\bf d}}

\begin{document}

\title{Development of Path Integral Monte Carlo Simulations with Localized Nodal Surfaces for Second-Row Elements}

\author{Burkhard Militzer} 
\affiliation{Department of Earth and Planetary Science, University of California, Berkeley}
\affiliation{Department of Astronomy, University of California, Berkeley}
\author{Kevin P. Driver} 
\affiliation{Department of Earth and Planetary Science, University of California, Berkeley}

\begin{abstract}
  We extend the applicability range of fermionic path integral Monte
  Carlo simulations to heavier elements and lower temperatures by
  introducing various localized nodal surfaces. Hartree-Fock nodes
  yield the most accurate prediction for pressure and internal energy
  that we combine with the results from density functional molecular
  dynamics simulations to obtain a consistent equation of state for
  hot, dense silicon under plasma conditions and in the regime of warm
  dense matter (2.3$-$18.6 g$\,$cm$^{-3}$, $5.0\,\times\,10^5 - 1.3\,
  \times\,10^8\,$K).  The shock Hugoniot curve is derived and the
  structure of the fluid is characterized with various pair
  correlation functions.
\end{abstract} 


\maketitle


The development of a first-principles methodology for warm dense
matter (WDM) applications that treats temperature effects consistently
is a key component of the stewardship of plasma
science~\cite{FESAC2009,HEDLP2009}. Indeed, technological progress in
high energy density physics (HEDP) applications, such as fusion
energy~\cite{Matzen2005, Cook2006}, shock-wave
physics~\cite{Cauble2001}, astrophysical
processes~\cite{VanHorn1991,Chabrier2002,Cotelo2011}, and
planetary~\cite{Fortney2009,Benuzzi2014} and stellar~\cite{Hansen1994}
interiors, relies on simulations for input and guidance.  WDM is
broadly described as the HEDP regime between condensed matter and
ideal plasmas, where strong electron correlation and quantum and
ionization effects are all important.

For the low temperature part of the WDM regime, density functional
molecular dynamics (DFT-MD) ~\cite{marx2000ab} is an accurate and
efficient first-principles simulation method. The thermal occupation
of electronic states is treated as a perturbation of the ground state
by Fermi-Dirac smearing~\cite{Gonze2001}. The main drawback of this
method is that it becomes computationally infeasible as electrons
occupy more bands with increasing temperature. Some alternative
DFT-MD-based methods, such as orbital-free DFT~\cite{Lambert2006,
  Lambert2007} and average-atom models~\cite{Rozsnyai2014}, have made
progress on overcoming the thermal-occupation deficiency, but efforts
to improve accuracy are still
underway~\cite{Karasiev2013,Sjostrom2014}.

Here, we focus on the development of the path integral Monte Carlo
(PIMC) method~\cite{Ce95}, which naturally incorporates finite
temperature quantum effects by working within the many-body thermal
density matrix formalism. The combination with Monte Carlo sampling
makes this approach one of the most appropriate first-principles
simulation techniques for quantum systems at finite temperature,
($T$). Since the length of the path scales like $1/T$, the method
becomes increasingly efficient for high temperatures.  Electrons and
nuclei are often treated equally as paths but here we treat the nuclei
classically because their zero-point motion is negligible for the
temperatures under consideration.

PIMC simulations with more than two electrons in a dense system suffer
from a fermion sign problem, which we solve by introducing the the
fixed node approximation~\cite{Ce91,Ce96} that restricts paths to
remain in the positive regions of a trial density matrix,
$\rho_{\scriptscriptstyle T}(\RR,\RR_t; t)>0$. The restricted path
integral reads,
\begin{equation}
\rho_F(\RR, \RR' ;\beta) =
\frac{1}{N!}\; \sum_\PP \; (-1)^\PP 
\! \! \! \! \! \! \! \! \! \! \! \! \! 
\int\limits_{
\begin{array}{c}
\scriptstyle
\RR \rightarrow \PP \RR', \, \rho_{\scriptscriptstyle T}>0
\end{array}
}
\! \! \! \! \! \! \! \! \! \! \! \! \! 
\dd\RR_t \;\; e^{-S[\RR_t] },
\label{restricted_PI} 
\end{equation}
where the action, $S$, weights every path and $\PP$ denotes
permutations of identical particles. The most common approximation to
the trial density matrix is a Slater determinant of single particle
density matrices,
\begin{equation}
\rho_T(\RR,\RRp;\beta)=\left|\left| \rho^{[1]}(r_{i},r'_{j};\beta) \right|\right|_{ij}\;,
\label{FP}
\eeq
in combination with the free particle (FP) density matrix,
\beq
\label{rho1}
\rho^{[1]}_0(r,r';\beta) = \sum_{k} e^{-\beta E_k} \, \Psi_k(r) \, \Psi_k^*(r')
\quad,
\end{equation}
derived from a sum over plane waves, $\Psi_k(r)$. The latter is
usually converted into Gaussian form~\cite{Ce91}. FP nodes becomes
exact in the limit of high temperature. Interaction effects have been
introduced to the nodal structure on the variational
level~\cite{MP00,MC00}.

In previous work~\cite{Mi01,Mi06,Mi09,Driver2012,Benedict2014,Driver2015}, we have
shown FP nodes can be sufficient to bridge the WDM regime for elements
as heavy as neon. FP nodes work for first-row elements because they
can still describe the occupation of the 1s state and DFT-MD works
well for lower temperatures where the second shell becomes occupied.
In order to simulate second-row elements with PIMC, one must go beyond
the FP nodal approximation and incorporate the effects of bound states
as we describe below in an application to silicon.

We chose to study silicon since it is a natural extension of our
original work on carbon and a prototype material with relevance in the
semiconductor industry~\cite{Liu2013}, geophysics and planetary
science~\cite{Benuzzi2014}, and
astrophysics~\cite{Herbst1989,Langer1990,MacKay1995,Schilke1997,Jones1994,Hansen1994}.
Silicon has a rich solid phase diagram, displaying 11 solid-state
phases under pressure, becoming metallic near 12
GPa~\cite{Mujica2003,Alfe2004,Hennig2010}. A number of dynamic shock
compression experiments have been
performed~\cite{Celliers1992,Alcon2004,Gilev1999,Gilev2004,Lower1998,Loveridge2001,Kishimura2010}.
Shock-compressed silicon has been studied theoretically with several
classical~\cite{Mogni2014,Reed2002,Li2014,Oleynik2006} and one DFT-MD
simulation~\cite{Swift2001} that investigated pressures up to 500 GPa
and temperatures up to $10^4$ K. Dynamical properties of shocked
silicon plasma states have also been studied extensively by
theoretical
approaches~\cite{Ng1995,Ng2002,Ng2011,Vorberger2010,Vorberger2012}.

We perform standard DFT-MD simulations using the VASP
code~\cite{VASP}. Exchange-correlation effects are described using the
Perdew-Burke-Ernzerhof~\cite{PBE} functional, which was not explicitly
designed for finite temperature~\cite{Brown2013}, but previous PIMC
and DFT-MD work~\cite{Mi01,Mi09,Driver2012,Driver2015}, has shown this
approximation is reliable. We use a plane-wave energy cut-off of up
4000 eV and a small-core ($r_{\rm core}$=1.0 \AA), PAW~\cite{PAW}
pseudopotential with 12 valence electrons. We used up to 9000 bands to
converge the thermal occupation to better than 10$^{-4}$. All
simulations are performed at the $\Gamma$-point of the Brillouin zone,
which is sufficient for high temperature fluids. 

Supercells with 8 atoms were used for $T \ge 2.5\,\times\,10^5\,$K
where the kinetic energy far outweighs the interaction energy, and
24-atoms were used at lower temperatures~\cite{Driver2015}. Additional
details are provided in the appendix. For the PIMC calculations, we
have used our own code CUPID~\cite{BM00}. The Coulomb interaction is
introduced through pair density matrices~\cite{Po88,Na95,MG06}. The
nodes are enforced at intervals of 1/8192 Ha, which means we need
between 4 and 2560 time slices for simulations in the temperature
range of $129-1\,\times\,10^6\,$K. It is sufficient to evaluate the
pair action only at intervals of 1/1024 Ha~\cite{MC00}.


We began our investigation of localized nodal approximations in PIMC
with the relatively simple, proof-of-concept problem of computing
internal energy and pressure of a stationary silicon atom (one nucleus
and 14 electrons) in a periodic cell over a wide temperature range. In
Fig.~\ref{atom}, we compared energies from DFT and PIMC using FP
nodes, where we found a discrepancy of 5.2 Ha/atom already at
$2\,\times\,10^6\,$K that increased to 12.6 Ha at
$5\,\times\,10^5\,$K. We attributed this discrepancy primarily to the
FP nodal approximation, which we have shown to work well only as long
as the second shell is not significantly occupied~\cite{Driver2012}.

\begin{figure}[htb]
\includegraphics[width=0.40\textwidth]{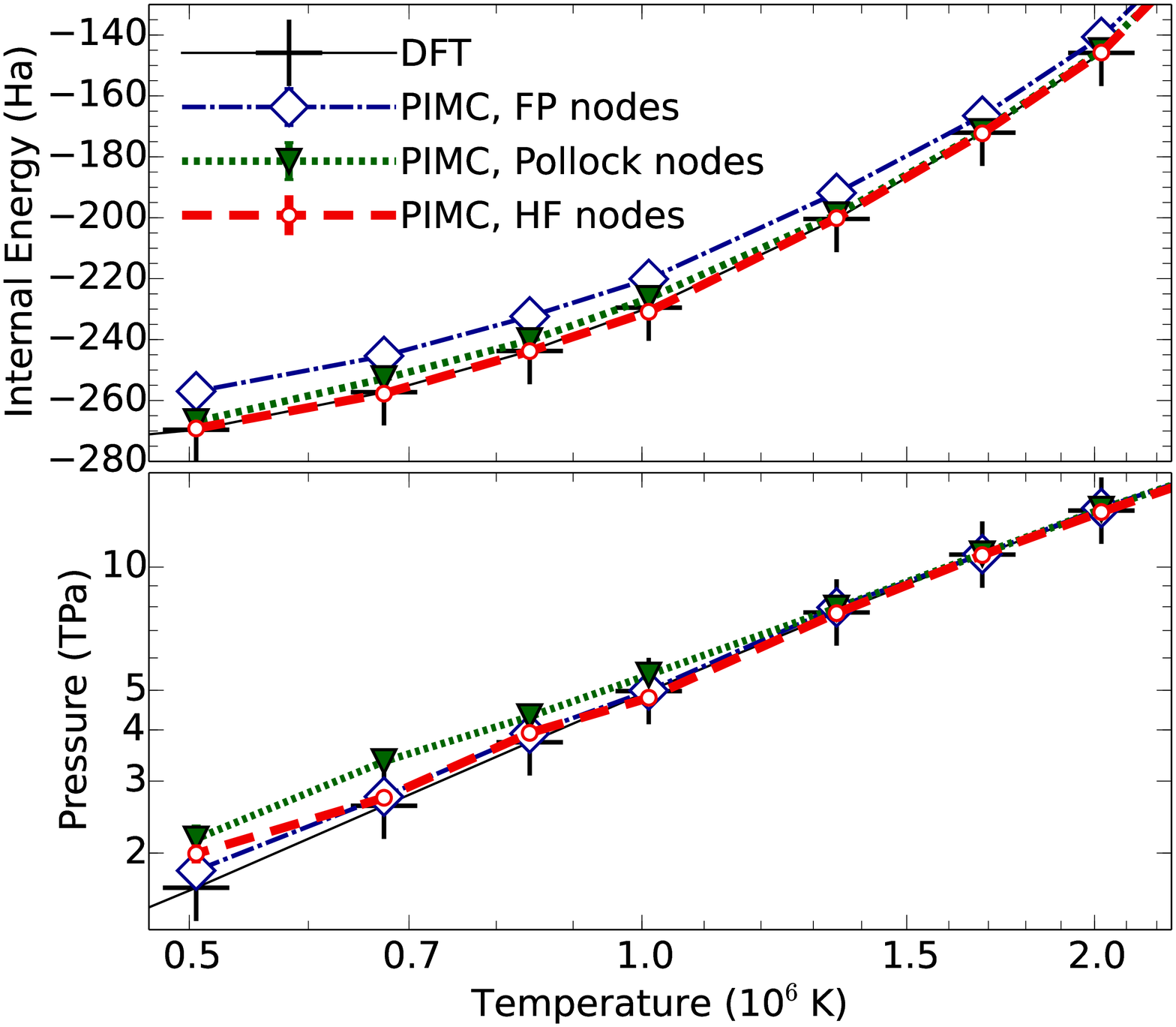}
\caption{Internal energy and pressure vs. temperature for a single silicon atom in periodic cell of 5.0 Bohr.}
\label{atom}
\end{figure}

We investigated two approaches to improve upon the FP nodal
approximation.  First, we added the bound eigenstates of the Coulomb
potential of the silicon nuclei, $\Psi_s(r-R_I)$, to the nodal
approximation in Eq.~\ref{rho1}:
\begin{equation}
  \rho^{[1]}(r,r',\beta) = \sum_{I=1}^{N} \sum_{s=0}^{n} e^{-\beta E_s} \Psi_s(r-R_I) \Psi_s^*(r'-R_I)\;,
  \label{pollock}
\end{equation}
where the number of states, $n$, needs to be at least 7 in each spin
channel in order to provide at least one bound state for every
electron. We used the efficient formulation of the Coulomb density
matrix put forth in Ref.~\cite{Po88} and hence refer to this
approximation as Pollock nodes. The 1s state ($n=1$) has been added to
PIMC nodes once before to simulate dense
hydrogen~\cite{ShumwayArchive}. However agreement with DFT predictions
and experimental results was not as good as expected because
additional approximations were introduced when the nodes were
enforced. Here we enforce the nodes strictly as outlined in
Refs.~\cite{Ce91,Ce96}.

The adoption of Pollock nodes reduced the energy deviation between DFT
and PIMC from 12.6 to 2.7 Ha at $5\,\times\,10^5\,$K. However, the
pressure deviations increased from 11 to 31\% (Fig.~\ref{atom}). We
tried to improve upon this result by varying the number of bound
states in Eq.~\ref{pollock}, testing different time steps, studying
various numbers of electrons, and finally by developing a
multi-determinental nodal surface in the spirit of quantum chemistry.
In the multi-determinantal approach, we adopted a sum of FP fermion
determinants where each is added to a different bound shell with the
appropriate $e^{-\beta E_s}$ weight. However, this approach did not
lead to a significant improvement in the predicted pressure.
This discrepancy led us to abandon the Pollock node approximation. We
concluded that the eigenstates of noninteracting particles in the
Coulomb potential are too confining for interacting electrons.

In our second approach, we constructed a thermal density matrix from
Hartree-Fock (HF) orbitals that we computed with the GAMESS
code~\cite{GAMESS} and expanded in a localized basis set (6-31++G). We
use again Eq.~\ref{pollock} but this time the functions $\Psi_s(r)$
become the HF orbitals, which are weighted by factors $e^{-\beta E_s}$
where $E_s$ is set to the corresponding HF eigenvalues. Our approach
differs from groundstate HF nodes~\cite{ShumwayCSSCMP2006}. With our
HF nodal approximation, we found perfect agreement with the DFT
prediction for the internal energy of the silicon atom over the entire
temperature range under consideration (Fig.~\ref{atom}). The resulting
PIMC energies are consistently lower than those obtained with other
two nodal approximations, which, as illustrate in the appendix,
implies a lower free energy~\cite{MC00} and establishes HF nodes as
the most accurate nodes among the three approximations considered
here.  The PIMC pressures derived with HF nodes agree within the 1
$\sigma$ error bars for all temperatures of $7\,\times\,10^5\,$K and
higher.  For $5\,\times\,10^5\,$K, a small pressure discrepancy
remained, but, given the large improvement over FP and Pollock nodes,
we decide to adopt HF nodes for our many-particle simulations with
moving nuclei that we discuss for the remainder of this article.

The evaluation of HF orbitals for many moving particles adds a
non-negligible burden to computation of the nodes. We vectorized this
part of the calculation by evaluating the orbitals for many positions
at once. We update the inverse of the determinants whenever possible
rather than recomputing it. Nevertheless, when one ion is moved, all
determinants need to be re-evaluated, which is not the case for FP
nodes that are independent of the ion positions. Despite this
additional cost, we were able to perform PIMC simulations with 8
nuclei and 112 electrons for temperatures of $1\,\times\,10^6\,$K
and above. 

\begin{figure}[htb]
\includegraphics[width=0.41\textwidth]{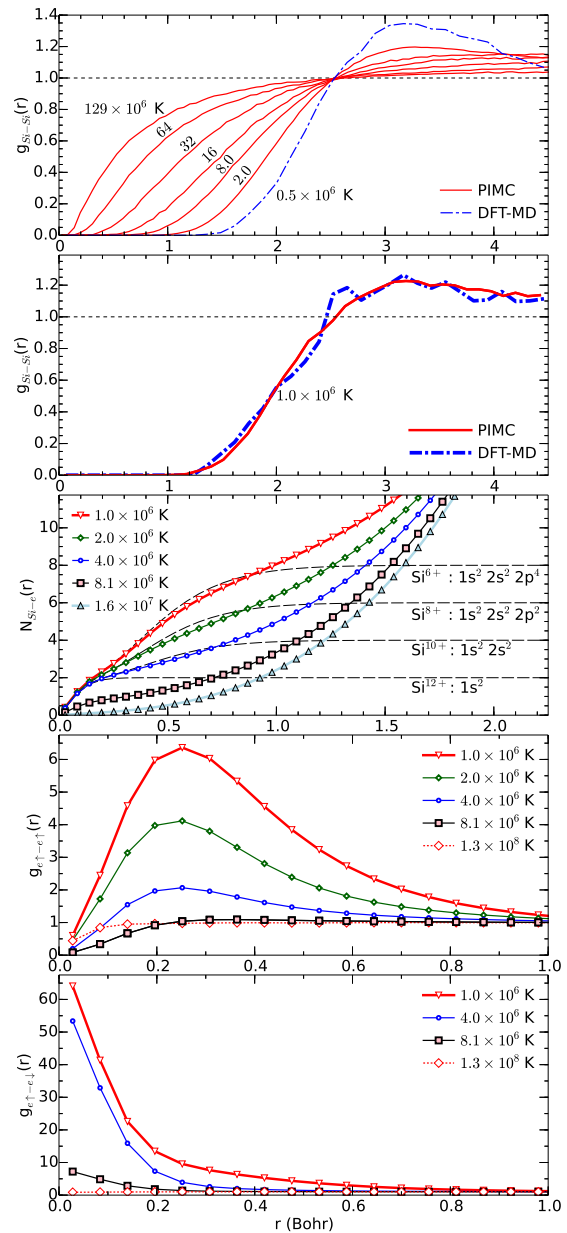}
\caption{The top two panels compare the nuclear pair correlation
  functions from PIMC and DFT-MD at various temperatures. The middle
  panel shows the integrated nucleus-electron pair correlation
  function, $N(r)$, computed with PIMC. Results are compared with an
  isolated ion in order to estimate the ionization state of the
  plasma.  The two lowest panels display the electron-electron pair
  correlation functions for pairs with parallel and opposite spins.
  All results are for 4-fold compression. }
\label{gr}
\end{figure}

We needed to introduce one more methodological development. Upon
introducing HF nodes into our simulations with moving nuclei, the
acceptance ratio for ion moves rapidly decayed to zero at lower and
intermediate temperatures as electron paths began to sample the bound
states at the nuclei. Because the nodal surfaces now depend on the
nuclear positions, node crossings are almost unavoidable when an ion
is moved. The crossing is almost exclusively triggered by nearby
electrons. The decay in efficiency was so detrimental that we could
not have obtained the smooth $g(r)$ functions in Fig.~\ref{gr} without
the development of multi-particle moves that relocate one nucleus and
nearby electrons at once. We needed to design an algorithm that
satisfies the detailed balance requirement~\cite{Ce95} and does not
rely on any permanent pairing of electrons and ions. We introduced a
localization function,
\begin{equation}
L_{Ij} = \int_0^\beta dt \, \left|\Psi_{1s}(r_j(t)-R_I)\right|^2\;,
\end{equation}
that assigns a probability of finding electron paths, $r_j(t)$, near
ion, $I$. Adopting concepts from the permutation sampling in
Ref.~\cite{Ce95}, we multiply these probabilities to construct a table
that contains all moves of one ion with up to four electron paths
including those that permute. Because $L_{Ij}$ is a very localized
function, the number of significant entries is fairly small so that
the table can be constructed efficiently. Once a particular move has
be selected from the table, we shift the entire group to a new
location within a box of 0.5 Bohr without otherwise changing their
paths. This leaves the function $L_{Ij}$ unchanged within the group,
which means detailed balance can be satisfied by adopting a
particularly simple expression for the acceptance ratio: the sum of
table entries for the new location divided by that for the original
coordinates. This procedure led to very efficient ion moves. To change
internal coordinates of electron paths, we keep relying the on single
and multi-electron moves~\cite{Ce95}.


\begin{figure}[htb]
\includegraphics[width=0.43\textwidth]{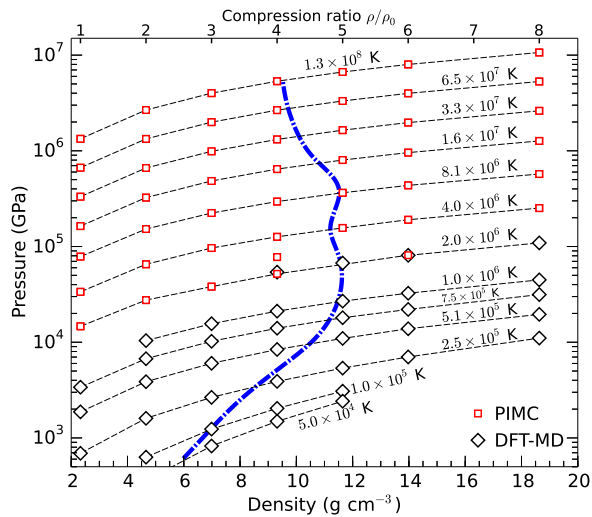}
\caption{ Pressure-density conditions of our PIMC and DFT-MD simulations. The blue line shows the shock Hugoniot curve. }
\label{hug}
\end{figure}

\begin{figure}[htb]
\includegraphics[width=0.43\textwidth]{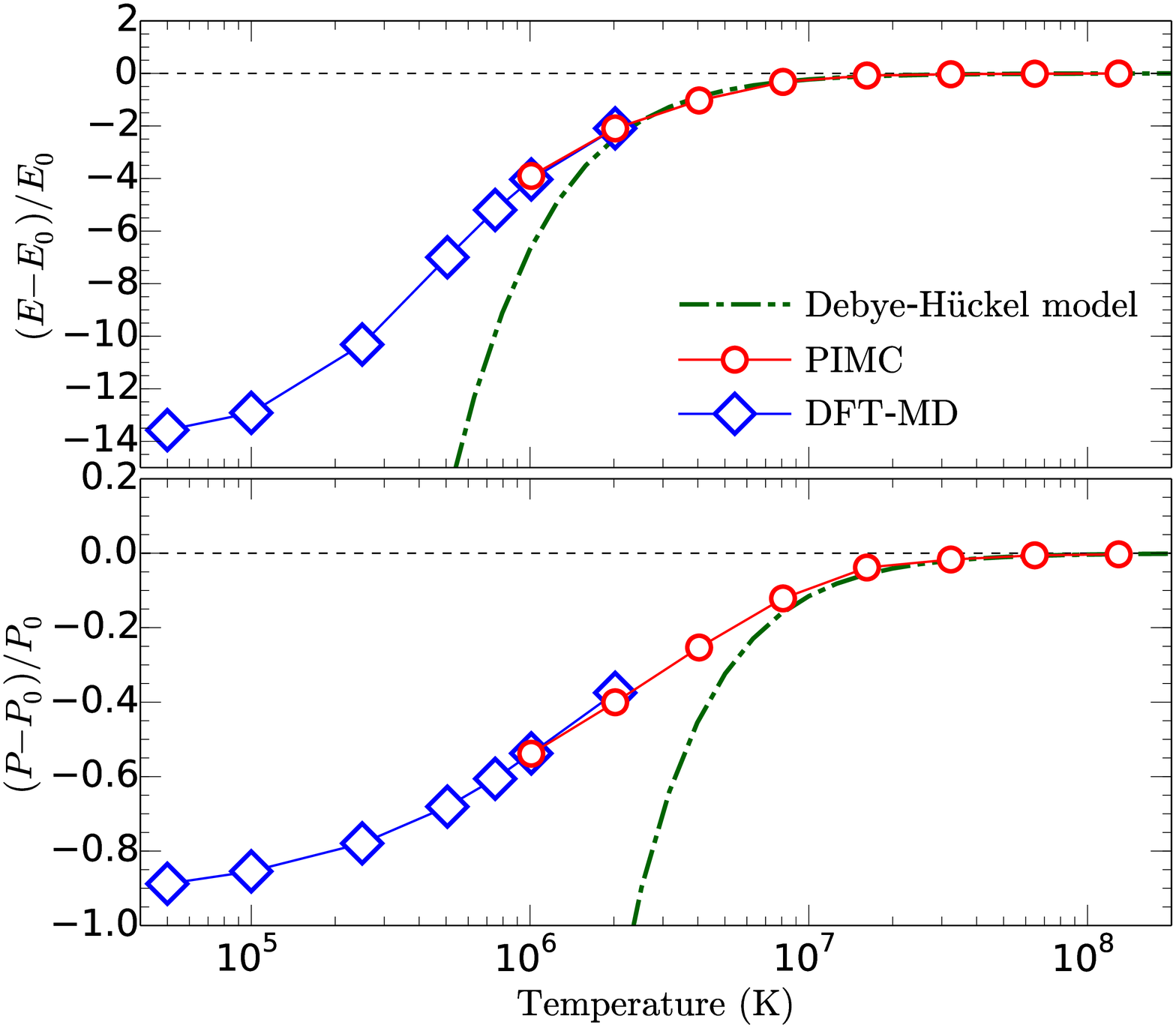}
\caption{ Internal energy and pressure for a silicon plasma at a
  density of 9.316 g$\,$cm$^{-3}$ are shown versus temperature. We
  plot the excess quantities relative to a fully ionized
  noninteracting plasma.}
\label{EP}
\end{figure}

Figure~\ref{hug} and Tab.~\ref{S1} summarize our equation of state
calculations.  For density interval of 1 to 8-fold the ambient density
of 2.329 g$\,$cm$^{-3}$, PIMC simulations with HF nodes were performed
for a temperature range of $129 - 2\,\times\,10^6\,$K and DFT-MD
simulations for $2\,-0.05\,\times\,10^6\,$K.  At $2\,\times\,10^6\,$K,
both methods yield consistent thermodynamic and structural properties
despite the fact that both techniques involve very different concepts
and approximations. The predicted internal energies deviate by up to 5
Ha/atom and the pressure by up to 4\%. A difference of 5 Ha/atom would
be equivalent to a 2.5\% difference in the ionization fraction of the
second shell. We attribute these deviations to a combined effect of
three approximations: the groundstate DFT exchange-correlation
functional, the frozen-core DFT pseudopotential, and our localized
nodes in PIMC. While it is difficult to disentangle the errors due to
these approximations, we anticipate that the discrepancies will be
reduced further when both methods are improved in the future.
Figure~\ref{EP} illustrates that the deviations between PIMC and
DFT-MD are small compared to the error in the Debye model. We only
plotted excess quantities relative to a fully ionized plasma model
because the total internal energy varies by over 10$\,$000 Ha/atom in
the parameter range of consideration.

Good agreement between PIMC and DFT-MD is found for the nuclear pair
correlation shown in Fig.~\ref{gr}. With PIMC we were also able to
derive the integrated nucleus-electron pair correlation function,
$N(r)$, that measures how many electron reside on average within a
radius, $r$, from a nucleus. Comparing the information at small $r$
with results for isolated ions, we can estimate the degree of
ionization in the plasma. For temperatures of 1, 2, and 4
$\times\,10^6\,$K, we estimate the average charge of the silicon ions
to be +6, +8, and +10 respectively. At higher temperature the 1s
states becomes partially ionized also.

The electron-electron pair correlation functions in Fig.~\ref{gr}
yield strong positive correlations, which underlines that multiple
electrons are bound to one nucleus. As the temperature is increased,
the positive correlation diminishes and eventually even the negative
correlations between electrons with parallel spins at small $r$ is
reduced.

Finally we derive the principal shock Hugoniot~\cite{Ze66}. Under
shock compression, a material changes from a initial state with
internal energy, pressure, and volume ($E_0=-289.166\,$Ha/atom,
$P_0$=1 bar, $V_0$ from $\rho_0=2.329$ g$\,$cm$^{-3}$) to a final
state denoted by $(E,P,V)$ that we can predict theoretically. The
shock compression ratio, $\rho/\rho_0$, is controlled by interaction
effects and by excitations of internal degrees of freedom. In
Fig.~\ref{hug}, a maximum compression ratio of 4.99 is reached for
1.6$\,\times\,10^6\,$K where approximately 7 of 14 electrons
have been ionized. A second compression maximum of 4.95 is predicted
to occur at 8.3$\,\times\,10^6\,$K, which is caused by the
ionization of the 1s state.  As we have seen for
neon~\cite{Driver2015}, the temperature is too high for this maximum
to be studied with DFT-MD. Therefore a combined PIMC and DFT-MD
approach is needed to study all features of the principal Hugoniot
curve.


By constructing a thermal density matrix with HF orbitals for the
purpose of computing fermion nodes, we were able to perform PIMC
simulations with heavier elements than was possible before.  Through
the optimized evaluation of such nodes and the adoption of
multi-particle Monte Carlo moves we were able to put together an
efficient algorithm and derive the equation of state of silicon
plasmas. At lower temperature, we add results from standard DFT-MD
simulations. By combining both techniques, we provide a
first-principles treatment for all second-row elements in the regime
of warm dense matter and for plasma conditions.

\acknowledgments{This research is supported by the U.S. Department of
  Energy, Grant No. DE-SC0010517. Computational support was provided
  by NASA, NERSC, and NCAR. B. Austin provided support for reading in
  the Gamess orbitals. J. Demmel gave advice on the numerics of
  determinant evaluations.}



\begin{thebibliography}{10}

\bibitem{FESAC2009}
R.~Betti, editor.
\newblock {\em Advancing the Science of High Energy Density Laboratory
  Plasmas}, Washington D.C., 2009. Office of Fusion Energy Science
  (OFES)/Fusion Energy Science Advisory Committee (FESAC).

\bibitem{HEDLP2009}
R.~Rosner, D.~Hammer, and T.~Rothman, editors.
\newblock {\em Basic Research Needs for high energy density laboratory
  physics}, Washington D.C., 2009. U.S. Department of Energy.

\bibitem{Matzen2005}
M.~Keith Matzen, M.~A. Sweeney, R.~G. Adams, J.~R. Asay, J.~E. Bailey, G.~R.\
  Bennett, D.~E. Bliss, D.~D. Bloomquist, T.~A. Brunner, R.~B. Campbell, G.~A.
  Chandler, C.~A\~. Coverdale, M.~E. Cuneo, J.-P. Davis, C.~Deeney, M.~P.
  Desjarlais, G.~L. Donovan, C.~J. Garasi, T.~A. Hail\~l, C.~A. Hall, D.~L.
  Hanson, M.~J. Hurst, B.~Jones, M.~D. Knudson, R.~J. Leeper, R.~W. Lem\~ke,
  M.~G. Mazarakis, D.~H. McDaniel, T.~A. Mehlhorn, T.~J. Nash, C.~L. Olson,
  J.\~L. Porter, P.~K. Rambo, S.~E. Rosenthal, G.~A. Rochau, L.~E. Ruggles,
  C.~L. Ruiz, J.~F. Sanford, T. W. L. a\ nd~Seamen, D.~B. Sinars, S.~A. Slutz,
  I.~C. Smith, K.~W. Struve, W.~A. Stygar, R.\~A. Vesey, E.~A. Weinbrecht,
  D.~F. Wenger, and E.~P. Yu.
\newblock Pulsed-power-driven high energy density physics and inertial
  confinement fusion research.
\newblock {\em Phys. Plasmas}, 12:055503, 2005.

\bibitem{Cook2006}
I.~Cook.
\newblock {Materials research for fusion energy}.
\newblock {\em Nat. Mater.}, 5:77, 2006.

\bibitem{Cauble2001}
R.~Cauble, D.~K. Bradley, P.~M. Celliers, G.~W. Collins, L.~B.~Da Siliva, and
  S.~J. Moon.
\newblock {Experiments Using Laser–driven Shockwaves for EOS and Transport
  Measurements}.
\newblock {\em Contrib. Plasma Physics}, 41:239, 2001.

\bibitem{VanHorn1991}
H.~M.~Van Horn.
\newblock {Dense astrophysical plasmas}.
\newblock {\em Science}, 252:384, 1991.

\bibitem{Chabrier2002}
G.~Chabrier, F.~Douchin, and A.~Y. Potekhin.
\newblock {Dense astrophysical plasmas}.
\newblock {\em J. Phys.: Condens. Matter}, 14:9133, 2002.

\bibitem{Cotelo2011}
M.~Cotelo, P.~Velarde adn A. G. de~la Varga, and C.~Garcia-Fernandez.
\newblock {\em Astrophys. Space Sci.}, 336:53, 2011.

\bibitem{Fortney2009}
J.~J. Fortney, S.~H. Glenzer, M.~Koenig, B.~Militzer, D.~Saumon, and
  D.~Valencia.
\newblock {Frontiers of the physics of dense plasmas and planetary interiors:
  Experiments, theory, and applications}.
\newblock {\em Phys. Plasmas}, 16:041003, 2009.

\bibitem{Benuzzi2014}
A.~Bennuzzi-Mounaix, S.~Mazevet, A.~Ravasio, T.~Vinci, A.~Denoeud, M.~Koenig,
  N.~Amadou, E.~Brambrink, F.~Festa, A.~Levy, M.~Harmand, S.~Brygoo, G.~Huser,
  V.~Recoules, J.~Bouchet, G.~Morard, F.~Guyot, T.~de~Resseguier, K.~Myanishi,
  N.~Ozaki, F.~Dorchies, J.~Gaudin, P.~M. Leguay, O.~Peyrusse, O.~Henry,
  D.~Raffestin, S.~Le Pape, R.~Smith, and R.~Musella.
\newblock {Progress in warm dense matter study with applications to
  planetology}.
\newblock {\em Phys. Scr.}, T161:014060, 2014.

\bibitem{Hansen1994}
C.J. Hansen and S.D. Kawaler.
\newblock {\em Stellar Interiors: Physical Principles, Structure, and
  Evolution}, volume~1 of {\em Astronomy and astrophysics library}.
\newblock Springer-Verlag New York, 1994.

\bibitem{marx2000ab}
Dominik Marx and Jurg Hutter.
\newblock Ab initio molecular dynamics: Theory and implementation.
\newblock {\em Modern methods and algorithms of quantum chemistry}, 1:301--449,
  2000.

\bibitem{Gonze2001}
M.~Verstraete and X.~Gonze.
\newblock {\em Phys. Rev. B}, 65, 2001.

\bibitem{Lambert2006}
{F. Lambert and J. Cl\'{e}rouin and S. Mazevet}.
\newblock {Structural and dynamical properties of hot dense matter by a
  Thomas-Fermi-Dirac molecular dynamics}.
\newblock {\em Europhys. Lett.}, 75:681, 2006.

\bibitem{Lambert2007}
F.~Lambert, J.~Cl{\'{e}}rouin, S.~Mazevet, and D.~Gilles.
\newblock Properties of hot dense plasmas by orbital-free molecular dynamics.
\newblock {\em Contrib. Plasma Phys.}, 47:272--280, 2007.

\bibitem{Rozsnyai2014}
B.~F. Rozsnyai.
\newblock Equation of state calculations based on the self-consistent
  ion-sphere and ion-correlation average atom models.
\newblock {\em High. Energy Dens. Phys.}, 16, 2014.

\bibitem{Karasiev2013}
V.~V. Karasiev, D.~Chakraborty, O.~A. Shukruto, and S.~B. Trickey.
\newblock {\em Phys. Rev. B}, 88, 2013.

\bibitem{Sjostrom2014}
T.~Sjostrom and J.~Daligault.
\newblock {\em Phys. Rev. Lett.}, 113, 2014.

\bibitem{Ce95}
D.~M. Ceperley.
\newblock {\em Rev. Mod. Phys.}, 67:279, 1995.

\bibitem{Ce91}
D.~M. Ceperley.
\newblock {\em J. Stat. Phys.}, 63:1237, 1991.

\bibitem{Ce96}
D.~M. Ceperley.
\newblock In Ed.~K. Binder and G.~Ciccotti, editors, {\em Monte Carlo and
  Molecular Dynamics of Condensed Matter Systems}. Editrice Compositori,
  Bologna, Italy, 1996.

\bibitem{MP00}
B.~Militzer and E.~L. Pollock.
\newblock {\em Phys. Rev. E}, 61:3470, 2000.

\bibitem{MC00}
B.~Militzer and D.~M. Ceperley.
\newblock {\em Phys. Rev. Lett.}, 85:1890, 2000.

\bibitem{Mi01}
B.~Militzer, D.~M. Ceperley, J.~D. Kress, J.~D. Johnson, L.~A. Collins, and
  S.~Mazevet.
\newblock {\em Phys. Rev. Lett.}, 87:275502, 2001.

\bibitem{Mi06}
B.~Militzer.
\newblock {\em Phys. Rev. Lett.}, 97:175501, 2006.

\bibitem{Mi09}
B.~Militzer.
\newblock {\em Phys. Rev. B}, 79:155105, 2009.

\bibitem{Driver2012}
K.~P. Driver and B.~Militzer.
\newblock {All-electron Path Integral Monte Carlo Simulations of Warm Dense
  Matter: Application to Water and Carbon Plasmas}.
\newblock {\em Phys. Rev. Lett.}, 108:115502, 2012.

\bibitem{Benedict2014}
L.~X. Benedict, K.~P. Driver, S.~Hamel, B.~Militzer, T.~Qi, A.~A. Correa,
  A.~Saul, and E.~Schwegler.
\newblock A multiphase equation of state for carbon addressing high pressures
  and temperatures.
\newblock {\em Phys. Rev. B}, 89:224109, 2014.

\bibitem{Driver2015}
K.~P. Driver and B.~Militzer.
\newblock {First-principles simulations and shock Hugoniot calculations of warm
  dense neon}.
\newblock {\em Phys. Rev. B}, 91:045103, 2015.

\bibitem{Liu2013}
{J. Liu and J. Cao and S. Kaierle}, editor.
\newblock {\em {Deformation behavior of single crystal silicon induced by laser
  shock peening}}, volume 8796. American Institute of Physics, 2013.

\bibitem{Herbst1989}
E.~Herbst, Tj~Millar, S.~Wlodek, and Dk~Bohme.
\newblock {The chemistry of silicon in dense interstellar clouds}.
\newblock {\em Astron. Astrophys.}, 222:205, 1989.

\bibitem{Langer1990}
William~D. Langer and a.~E. Glassgold.
\newblock {Silicon chemistry in interstellar clouds}.
\newblock {\em Astrophys. J.}, 352:123, 1990.

\bibitem{MacKay1995}
D.~D.~S. MacKay.
\newblock {\em Mon. Not. R. Astron. Soc.}, 274:694, 1995.

\bibitem{Schilke1997}
P~Schilke, C.\~{}M. Walmsley, G~{Pineau des Forets}, and D.\~{}R. Flower.
\newblock {SiO production in interstellar shocks.}
\newblock {\em Astron. Astrophys.}, 321:293, 1997.

\bibitem{Jones1994}
A.~P. Jones, A.~G. G.~M. Tielens, D.~J. Hollenbach, and C.~F. Mckee.
\newblock {Grain destruction in shocks in the interstellar medium}.
\newblock {\em Astrophys. J.}, 433:797, 1994.

\bibitem{Mujica2003}
{A. Mujica and A. Rubio and A. Mu\~{n}oz}.
\newblock {High-pressure phases of group-IV , III – V , and II – VI
  compounds}.
\newblock {\em Rev. Mod. Phys.}, 75:863, 2003.

\bibitem{Alfe2004}
D.~Alf\`{e}, M.~J. Gillan, M.~D. Towler, and R.~J. Needs.
\newblock {Diamond and $\beta$-tin structures of Si studied with quantum Monte
  Carlo calculations}.
\newblock {\em Phys. Rev. B}, 70:214102, 2004.

\bibitem{Hennig2010}
R.~G. Hennig, A.~Wadehra, K.~P. Driver, W.~D. Parker, C.~J. Umrigar, and J.~W.
  Wilkins.
\newblock {Phase transformation in Si from semiconducting diamond to metallic
  beta-Sn phase in QMC and DFT under hydrostatic and anisotropic stress}.
\newblock {\em Phys. Rev. B}, 82:014101, 2010.

\bibitem{Celliers1992}
P.~Celliers, a.~Ng, G.~Xu, and a.~Forsman.
\newblock {Thermal equilibration in a shock wave}.
\newblock {\em Phys. Rev. Lett.}, 68:2305, 1992.

\bibitem{Alcon2004}
{R. R. Alcon and D. L. Robbins and S. A. Sheffield and D. B. Stahl and J. N.
  Fritz}.
\newblock {Shock Compression of Silicon Polymer Foams with a Range of Initial
  Densities}.
\newblock {\em AIP Conf. Proc.}, 706:651, 1977.

\bibitem{Gilev1999}
Sd~Gilev and Am~Trubachev.
\newblock {Metallization of Monocrystalline Silicon under Shock Compression}.
\newblock {\em Phys. Status Solidi B Basic Res.}, 211:379, 1999.

\bibitem{Gilev2004}
S~D Gilev and a~M Trubachev.
\newblock {Metallization of silicon in a shock wave: the metallization
  threshold and ultrahigh defect densities}.
\newblock {\em J. Phys. Condens. Matter}, 16(46):8139, 2004.

\bibitem{Lower1998}
Th. L\"{o}wer, V.~Kondrashov, M.~Basko, a.~Kendl, J.~Meyer-ter Vehn, R.~Sigel,
  and a.~Ng.
\newblock {Reflectivity and Optical Brightness of Laser-Induced Shocks in
  Silicon}.
\newblock {\em Phys. Rev. Lett.}, 80:4000, 1998.

\bibitem{Loveridge2001}
a.~Loveridge-Smith, a.~Allen, J.~Belak, T.~Boehly, a.~Hauer, B.~Holian,
  D.~Kalantar, G.~Kyrala, R.~W. Lee, P.~Lomdahl, M.~a. Meyers, D.~Paisley,
  S.~Pollaine, B.~Remington, D.~C. Swift, S.~Weber, and J.~S. Wark.
\newblock {Anomalous elastic response of silicon to uniaxial shock compression
  on nanosecond time scales}.
\newblock {\em Phys. Rev. Lett.}, 86:2349, 2001.

\bibitem{Kishimura2010}
H~Kishimura, H~Matsumoto, and N~N Thadhani.
\newblock {Effect of shock compression on single crystalline silicon}.
\newblock {\em J. Phys. Conf. Ser.}, 215:012145, 2010.

\bibitem{Mogni2014}
Gabriele Mogni, Andrew Higginbotham, Katalin Ga\'{a}l-Nagy, Nigel Park, and
  Justin~S. Wark.
\newblock {Molecular dynamics simulations of shock-compressed single-crystal
  silicon}.
\newblock {\em Phys. Rev. B}, 89:064104, 2014.

\bibitem{Reed2002}
Evan~J. Reed.
\newblock {Hugoniot Constraint Molecular Dynamics Study of a Transformation to
  a Metastable Phase in Shocked Silicon}.
\newblock {\em AIP Conf. Proc.}, 620:343, 2002.

\bibitem{Li2014}
Zhongyu Li, Di~Chen, Jing Wang, and Lin Shao.
\newblock {Molecular dynamics simulation of Coulomb explosion, melting and
  shock wave creation in silicon after an ionization pulse}.
\newblock {\em J. Appl. Phys.}, 115:143507, 2014.

\bibitem{Oleynik2006}
I.~I. Oleynik, S.~V. Zybin, M.~L. Elert, and C.~T. White.
\newblock {Nanoscale molecular dynamics simulaton of shock compression of
  silicon}.
\newblock {\em AIP Conf. Proc.}, 845(2006):413, 2006.

\bibitem{Swift2001}
D.~Swift, G.~Ackland, a.~Hauer, and G.~Kyrala.
\newblock {First-principles equations of state for simulations of shock waves
  in silicon}.
\newblock {\em Phys. Rev. B}, 64:214107, 2001.

\bibitem{Ng1995}
A.~Ng, P.~Celliers, G.~Xu, and A.~Forsman.
\newblock {Electron-ion equilibration in a strongly coupled plasma}.
\newblock {\em Phys. Rev. E}, 52:4299, 1995.

\bibitem{Ng2002}
M.~D. Furnish, N.~N. Thadhani, and Y.~Horie, editors.
\newblock {\em Shock Waves and Plasma Physics}. American Institute of Physics,
  2002.

\bibitem{Ng2011}
A.~Ng.
\newblock {\em International Journal of Quantum Chemistry}, 112:150, 2011.

\bibitem{Vorberger2010}
J.~Vorberger, D.~O. Gericke, Th~Bornath, and M.~Schlanges.
\newblock {Energy relaxation in dense, strongly coupled two-temperature
  plasmas}.
\newblock {\em Phys. Rev. E}, 81:046404, 2010.

\bibitem{Vorberger2012}
J.~Vorberger, Z.~Donko, I.~M. Tkachenko, and D.~O. Gericke.
\newblock {Dynamic ion structure factor of warm dense matter}.
\newblock {\em Phys. Rev. Lett.}, 109:225001, 2012.

\bibitem{VASP}
G.~Kresse and J.~Furthm\"{u}ller.
\newblock {Efficient iterative schemes for ab initio total-energy calculations
  using a plane-wave basis set}.
\newblock {\em Phys. Rev. B}, 54:11169, 1996.

\bibitem{PBE}
J.~P. Perdew, K.~Burke, and M.~Ernzerhof.
\newblock {Generalized gradient approximation made simple}.
\newblock {\em Phys. Rev. Lett.}, 77:3865, 1996.

\bibitem{Brown2013}
E.~W. Brown, B.~K. Clark, J.~L. DuBois, and D.~M. Ceperley.
\newblock {Path-Integral Monte Carlo Simulations of the Warm Dense Homogeneous
  Electron Gas}.
\newblock {\em Phys. Rev. Lett.}, 110:146405, 2013.

\bibitem{PAW}
P.~E. Bl\"{o}chl.
\newblock {Projector augmented-wave method}.
\newblock {\em Phys. Rev. B}, 50:17953, 1994.

\bibitem{BM00}
B.~Militzer.
\newblock PhD thesis, University of Illinois at Urbana-Champaign, 2000.

\bibitem{Po88}
E.~L. Pollock.
\newblock {\em Comp. Phys. Comm.}, {52 }:49, 1988.

\bibitem{Na95}
V.~Natoli and D.~M. Ceperley.
\newblock {\em J. Comp. Phys.}, 117:171--178, 1995.

\bibitem{MG06}
B.~Militzer and R.~L. Graham.
\newblock {\em Journal of Physics and Chemistry of Solids}, 67:2136, 2006.

\bibitem{ShumwayArchive}
S. A. Khairallah and J. Shumway and E. Draeger, available on arXiv:1108.1711.

\bibitem{GAMESS}
M.W.Schmidt, K.K.Baldridge, J.A.Boatz, S.T.Elbert, M.S.Gordon, J.J.Jensen,
  S.Koseki, N.Matsunaga, K.A.Nguyen, S.Su, T.L.Windus, M.Dupuis, and
  J.A.Montgomery.
\newblock {\em J.~Comput.~Chem.}, 14:1347, 1993.

\bibitem{ShumwayCSSCMP2006}
{J. Shumway}.
\newblock {All-Electron Path Integral Monte Carlo Simulations of Small Atoms
  and Molecules}.
\newblock In {\em Computer Simulation Studies in Condensed-Matter Physics
  {XVI}}, page 181. Springer Berlin Heidelberg, 2006.

\bibitem{Ze66}
Y.~B. Zeldovich and Y.~P. Raizer.
\newblock {\em Elements of Gasdynamics and the Classical Theory of Shock
  Waves}.
\newblock Academic Press, New York, 1968.

\end{thebibliography}

\subsection{Appendix: Internal Energy Comparison for Different Nodal Surface}

\begin{figure}[htb]
\renewcommand\thefigure{S1}
\includegraphics[width=0.40\textwidth]{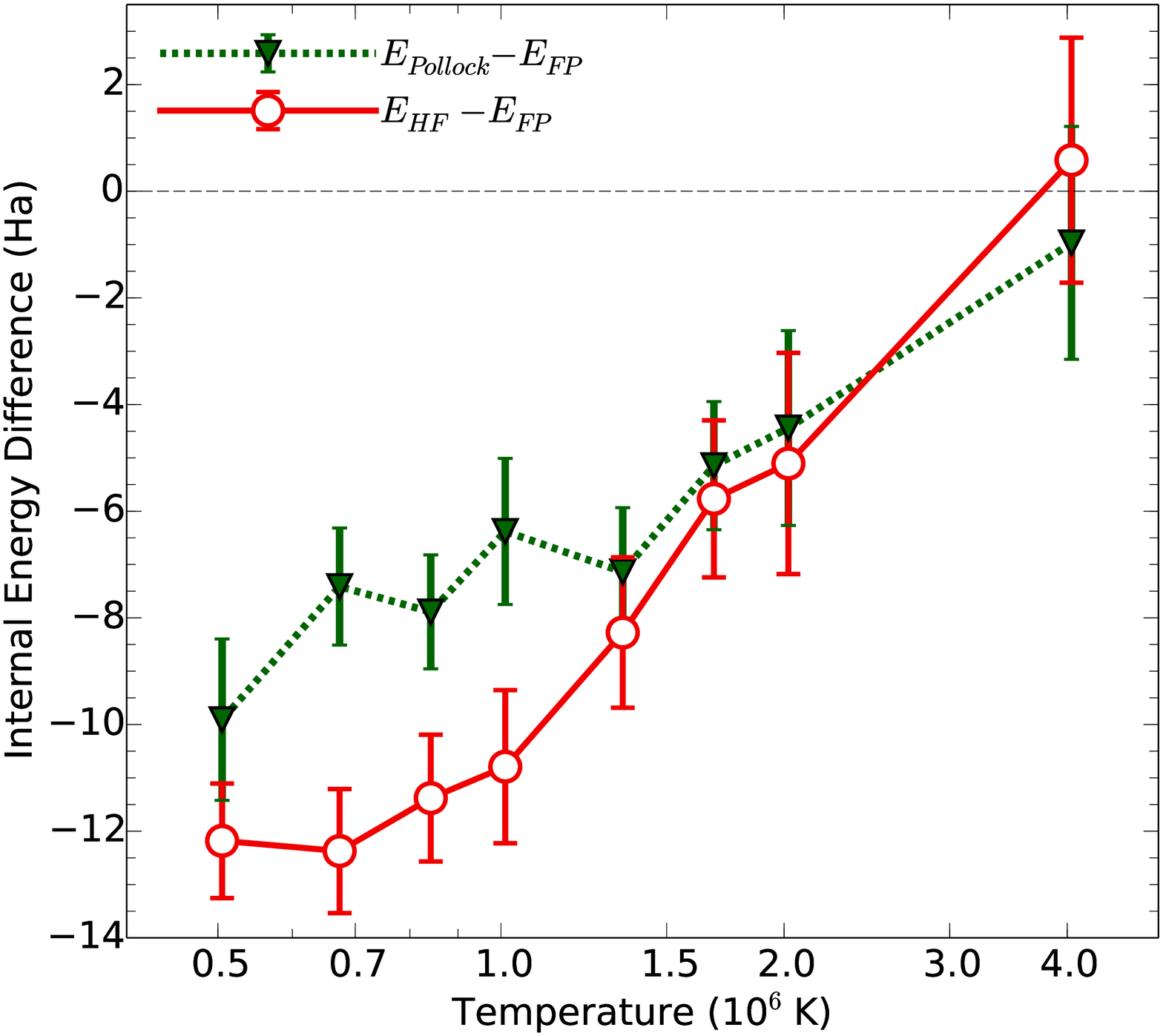}
\caption{Difference in internal energy between PIMC calculations using
  Pollock, Hartree-Fock (HF), and free-particle (FP) nodes for a
  single silicon atom in periodic cell of 5.0 Bohr.}
\label{atom}
\end{figure}

In Fig.~\ref{atom}, we plot the internal energy difference between
PIMC calculations with various nodal surfaces for the isolated silicon
atom that we already reported in a alternate format in figure 1 of the
article. For this specific example, one finds that FP particle nodes
are sufficiently accurate for $T \ge 4.0 \times 10^6\,$K. For $T \ge
1.7 \times 10^6\,$K, PIMC results obtained with Pollock nodes agree
reasonably well with predictions from HF nodes, our most reliable
nodal approximation.

The internal energies, $E$, obtained with HF nodes are consistently
lower than those with both other nodal surfaces. Therefore, the HF
nodes are our most accurate nodal surface because they lead to the
lowest free energy, $F$. In fermionic PIMC, the best nodal surface can
be established by minimizing the free energy~\cite{MC00}. $F$ is given
by the integral over the internal energy, $E$,
\begin{equation}
\beta_2 \, F(\beta_2,V) = \beta_1 \, F(\beta_1,V) + \int_{\beta_1}^{\beta_2} d \beta \, E(\beta,V)\;,
\end{equation}
where $\beta = 1/k_bT$. As starting point for the integration, we can
take the classical, high temperature limit ($\beta_1 \to 0$) where the
predictions from all nodal surfaces agree. From Fig.~\ref{atom}, we
can conclude that our HF nodes are significantly better than FP and
Pollock nodes.
\\
~\\
~\\
~\\
~\\
~\\
~\\
~\\
~\\
~\\
~\\
~\\
\\

\onecolumngrid
\appendix


\begin{table}[h]
\label{S1}
\begin{tabular} {l c c r r c r r}
\hline
\hline
Method & $\rho$ (g$\,$cm$^{-3}$) & $\rho/\rho_0$  & $T (K)$~~~ & $E$ (Ha/atom) & $\epsilon_E$ (Ha/atom) & ~~~$P$ (GPa)~~~ & ~~~$\epsilon_P$ (GPa) \\
\hline
\hline
PIMC     &  2.329 & 1   & 129341301   & 9181.8930 & 3.6180   &   1335190.7 &   526.0 \\
PIMC     &  2.329 & 1   &  64670651   & 4574.1180 & 3.4980   &    666626.4 &   507.1 \\
PIMC     &  2.329 & 1   &  32335325   & 2265.3520 & 3.2550   &    331901.2 &   470.4 \\
PIMC     &  2.329 & 1   &  16167663   & 1100.3200 & 1.8800   &    163533.0 &   271.7 \\
PIMC     &  2.329 & 1   &   8083831   &  482.3270 & 1.4570   &     78489.3 &   203.0 \\
PIMC     &  2.329 & 1   &   4041916   &   35.6380 & 1.0560   &     33636.6 &   151.5 \\
PIMC     &  2.329 & 1   &   2020958   & $-$128.4320 & 1.4580   &     14663.8 &   211.6 \\
DFT-MD   &  2.329 & 1   &    750000   & $-$247.1570 & 0.0008   &      3386.9 &     0.1 \\
DFT-MD   &  2.329 & 1   &    505239   & $-$266.9010 & 0.0006   &      1872.2 &     0.1 \\
DFT-MD   &  2.329 & 1   &    250000   & $-$282.8030 & 0.0005   &       691.0 &     0.4 \\
\hline
PIMC     &  4.658 & 2   & 129341301   & 9171.0380 & 3.4810   &   2668660.3 &  1008.1 \\
PIMC     &  4.658 & 2   &  64670651   & 4563.5600 & 3.3340   &   1331789.0 &   969.2 \\
PIMC     &  4.658 & 2   &  32335325   & 2251.5260 & 3.3350   &    661773.8 &   965.2 \\
PIMC     &  4.658 & 2   &  16167663   & 1080.9040 & 1.7860   &    324383.1 &   514.1 \\
PIMC     &  4.658 & 2   &   8083831   &  439.5750 & 1.6430   &    152336.7 &   445.7 \\
PIMC     &  4.658 & 2   &   4041916   &   12.3270 & 1.1950   &     65200.4 &   346.3 \\
PIMC     &  4.658 & 2   &   2020958   & $-$146.2080 & 1.3640   &     27553.5 &   396.8 \\
DFT-MD   &  4.658 & 2   &   1010479   & $-$231.9070 & 0.0030   &     10398.8 &     5.3 \\
DFT-MD   &  4.658 & 2   &    750000   & $-$252.4430 & 0.0030   &      6724.3 &     4.3 \\
DFT-MD   &  4.658 & 2   &    505239   & $-$269.6440 & 0.0040   &      3866.4 &     8.6 \\
DFT-MD   &  4.658 & 2   &    250000   & $-$283.2340 & 0.0060   &      1602.7 &     5.8 \\
DFT-MD   &  4.658 & 2   &    100000   & $-$287.4370 & 0.0080   &       629.7 &     4.1 \\
DFT-MD   &  4.658 & 2   &     50000   & $-$288.4070 & 0.0060   &       358.8 &     4.3 \\
\hline
PIMC     &  6.987 & 3   & 129341301   & 9169.1970 & 3.1060   &   4003930.0 &  1348.8 \\
PIMC     &  6.987 & 3   &  64670651   & 4547.5440 & 3.5220   &   1992509.2 &  1532.7 \\
PIMC     &  6.987 & 3   &  32335325   & 2235.4160 & 3.2940   &    987707.6 &  1428.8 \\
PIMC     &  6.987 & 3   &  16167663   & 1067.9070 & 2.2080   &    483879.2 &   956.5 \\
PIMC     &  6.987 & 3   &   8083831   &  408.9970 & 1.9730   &    223935.3 &   835.2 \\
PIMC     &  6.987 & 3   &   4041916   &    0.8250 & 1.5560   &     96497.2 &   677.6 \\
PIMC     &  6.987 & 3   &   2020958   & $-$159.5200 & 1.8080   &     38095.5 &   789.1 \\
DFT-MD   &  6.987 & 3   &   1010479   & $-$236.1540 & 0.0060   &     15607.2 &    27.4 \\
DFT-MD   &  6.987 & 3   &    750000   & $-$255.1520 & 0.0020   &     10228.1 &    21.0 \\
DFT-MD   &  6.987 & 3   &    505239   & $-$270.9450 & 0.0150   &      6020.6 &    27.0 \\
DFT-MD   &  6.987 & 3   &    250000   & $-$283.3580 & 0.0130   &      2653.2 &     7.8 \\
DFT-MD   &  6.987 & 3   &    100000   & $-$287.2920 & 0.0070   &      1237.0 &     5.8 \\
DFT-MD   &  6.987 & 3   &     50000   & $-$288.2310 & 0.0090   &       823.5 &     7.8 \\
\hline
PIMC     &  9.316 & 4   & 129341301   & 9160.5960 & 3.4840   &   5335104.8 &  2020.1 \\
PIMC     &  9.316 & 4   &  64670651   & 4551.1630 & 3.5780   &   2660714.0 &  2077.3 \\
\hline
\hline
\end{tabular}
\\
Continued on the following page.~~~~~~~~~~~~~~~~~~~~~~~~~~~~~~~~~~~~~~~~~~~~~~~~~~~~~~~~~~~~~~~~~~~~~~~~~~~~~~~~~~~~~~~~~~~~~~~~~~~~
\end{table}

\begin{table}[h]
\renewcommand\thetable{S1}
\begin{tabular} {l c c r r c r r}
\hline
\hline
Method & $\rho$ (g$\,$cm$^{-3}$) & $\rho/\rho_0$  & $T (K)$~~~ & $E$ (Ha/atom) & $\epsilon_E$ (Ha/atom) & ~~~$P$ (GPa)~~~ & ~~~$\epsilon_P$ (GPa) \\
\hline
\hline
PIMC     &  9.316 & 4   &  32335325   & 2228.9250 & 3.4630   &   1315670.2 &  2006.9 \\
PIMC     &  9.316 & 4   &  16167663   & 1059.1460 & 2.4510   &    644388.0 &  1408.0 \\
PIMC     &  9.316 & 4   &   8083831   &  390.6560 & 1.7500   &    295039.8 &   953.0 \\
PIMC     &  9.316 & 4   &   4041916   &   $-$9.0920 & 1.2470   &    126325.6 &   723.0 \\
PIMC     &  9.316 & 4   &   2694610   & -109.3620 & 1.6440   &     77593.7 &   954.1 \\
PIMC     &  9.316 & 4   &   2020958   & $-$162.7230 & 0.8820   &     51815.1 &   518.4 \\
DFT-MD   &  9.316 & 4   &   2020958   & $-$162.3860 & 0.0460   &     53994.0 &    49.3 \\
DFT-MD   &  9.316 & 4   &   1010479   & $-$238.9270 & 0.0090   &     21127.6 &    48.5 \\
DFT-MD   &  9.316 & 4   &    750000   & $-$256.8640 & 0.0140   &     14000.3 &    44.5 \\
DFT-MD   &  9.316 & 4   &    505239   & $-$271.6910 & 0.0090   &      8399.0 &    19.8 \\
DFT-MD   &  9.316 & 4   &    250000   & $-$283.3240 & 0.0080   &      3897.3 &     7.7 \\
DFT-MD   &  9.316 & 4   &    100000   & $-$287.0760 & 0.0100   &      2035.8 &     9.7 \\
DFT-MD   &  9.316 & 4   &     50000   & $-$287.9990 & 0.0060   &      1491.6 &     7.2 \\
\hline
PIMC     & 11.645 & 5   & 129341301   & 9154.7710 & 1.6460   &   6666339.6 &  1194.9 \\
PIMC     & 11.645 & 5   &  64670651   & 4536.3090 & 1.6700   &   3317133.6 &  1210.4 \\
PIMC     & 11.645 & 5   &  32335325   & 2224.0050 & 1.6780   &   1643487.2 &  1214.9 \\
PIMC     & 11.645 & 5   &  16167663   & 1047.8130 & 1.8210   &    800998.2 &  1306.9 \\
PIMC     & 11.645 & 5   &   8083831   &  371.3900 & 1.6760   &    364447.7 &  1143.3 \\
PIMC     & 11.645 & 5   &   4041916   &  $-$16.3540 & 1.3620   &    156361.8 &   993.2 \\
DFT-MD   & 11.645 & 5   &   2020958   & $-$166.7740 & 0.0710   &     67206.6 &    97.1 \\
DFT-MD   & 11.645 & 5   &   1010479   & $-$240.8560 & 0.0270   &     26892.2 &   107.7 \\
DFT-MD   & 11.645 & 5   &    750000   & $-$258.0590 & 0.0090   &     18026.4 &    43.7 \\
DFT-MD   & 11.645 & 5   &    505239   & $-$272.1630 & 0.0100   &     10955.4 &    23.8 \\
DFT-MD   & 11.645 & 5   &    250000   & $-$283.1680 & 0.0130   &      5363.5 &    12.8 \\
DFT-MD   & 11.645 & 5   &    100000   & $-$286.7950 & 0.0070   &      3078.9 &     7.0 \\
DFT-MD   & 11.645 & 5   &     50000   & $-$287.6990 & 0.0050   &      2420.6 &     6.6 \\
\hline
PIMC     & 13.974 & 6   & 129341301   & 9144.6710 & 3.5480   &   7992810.7 &  3087.8 \\
PIMC     & 13.974 & 6   &  64670651   & 4540.0170 & 3.3760   &   3986259.3 &  2943.9 \\
PIMC     & 13.974 & 6   &  32335325   & 2220.7620 & 3.4170   &   1972368.7 &  2960.8 \\
PIMC     & 13.974 & 6   &  16167663   & 1038.4910 & 2.4970   &    957890.7 &  2124.0 \\
PIMC     & 13.974 & 6   &   8083831   &  361.2450 & 2.2060   &    435422.9 &  1865.3 \\
PIMC     & 13.974 & 6   &   4041916   &  $-$17.6870 & 1.9410   &    190435.9 &  1688.3 \\
PIMC     & 13.974 & 6   &   2020958   & $-$164.5430 & 1.4990   &     81182.1 &  1308.2 \\
DFT-MD   & 13.974 & 6   &   2020958   & $-$169.9650 & 0.1390   &     80923.8 &   182.8 \\
DFT-MD   & 13.974 & 6   &   1010479   & $-$242.3660 & 0.0470   &     32508.1 &   136.8 \\
DFT-MD   & 13.974 & 6   &    750000   & $-$258.9620 & 0.0180   &     22007.2 &    80.1 \\
DFT-MD   & 13.974 & 6   &    505239   & $-$272.4350 & 0.0370   &     13834.1 &    93.4 \\
DFT-MD   & 13.974 & 6   &    250000   & $-$282.9850 & 0.0160   &      7014.2 &    18.5 \\
\hline
PIMC     & 18.632 & 8   & 129341301   & 9137.0560 & 3.4650   &  10652592.7 &  4018.6 \\
PIMC     & 18.632 & 8   &  64670651   & 4523.4160 & 3.3800   &   5300935.9 &  3908.0 \\
PIMC     & 18.632 & 8   &  32335325   & 2205.0500 & 3.5490   &   2617832.7 &  4099.8 \\
PIMC     & 18.632 & 8   &  16167663   & 1022.1250 & 1.9640   &   1268337.8 &  2234.7 \\
PIMC     & 18.632 & 8   &   8083831   &  340.2690 & 1.8940   &    572026.0 &  2134.0 \\
PIMC     & 18.632 & 8   &   4041916   &  $-$25.8850 & 1.3170   &    252032.9 &  1535.4 \\
DFT-MD   & 18.632 & 8   &   2020958   & $-$175.0180 & 0.0770   &    109041.1 &   174.0 \\
DFT-MD   & 18.632 & 8   &   1010479   & $-$244.5310 & 0.0180   &     44810.9 &   106.4 \\
DFT-MD   & 18.632 & 8   &    750000   & $-$260.0340 & 0.0490   &     31467.4 &   246.7 \\
DFT-MD   & 18.632 & 8   &    505239   & $-$272.7980 & 0.0060   &     19547.6 &    32.5 \\
DFT-MD   & 18.632 & 8   &    250000   & $-$282.4490 & 0.0080   &     11047.4 &    16.9 \\
\hline
\hline
\end{tabular}
\caption{Equation of state table for hot, dense silicon providing the internal energy, $E$, and pressure, $P$, as function of density, $\rho$, and temperature, $T$. $\epsilon$ denotes the statistical error bars. The zero of energy taken from a completely ionized system. $\rho_0=$ 2.329 g$\,$cm$^{-3}$.}
\label{table1}
\end{table}

\end{document}